\documentclass[aps,prev,twocolumn,preprintnumbers,floatfix,nofootinbib]{revtex4-1}
\pdfoutput=1 % if your are submitting a pdflatex (i.e. if you have
\usepackage{amsmath}
\usepackage{amsfonts}
\usepackage{amssymb}
\usepackage{graphicx}
\usepackage{bm}
\usepackage{times}
\usepackage{hyperref}
\usepackage{xfrac}
\usepackage{slashed}
\usepackage{color}
\usepackage{subfigure}
\usepackage{threeparttable}

\begin{document}

\preprint{IIPDM-2019}

\title{\boldmath Secluded Dark Matter in light of the Cherenkov Telescope Array (CTA)}

\author{Clarissa Siqueira}
\email{csiqueira@iip.ufrn.br}

\affiliation{International Institute of Physics, Universidade Federal do Rio Grande do Norte,
Campus Universit\'ario, Lagoa Nova, Natal-RN 59078-970, Brazil}

\begin{abstract}
%The extensive search for particle Dark Matter remains one of the most important challenges in which concerns particle physics and cosmology. In this work, we intend to show how Secluded models are and will be constrained in light of the most recent and prospects of indirect gamma-ray experiments. In order to do this, we choose two different scenarios, the first one choosing the mediator as a Higgs-like particle differing from the standard one by the mass, in the second case, we take the mediator as a particle whose decays predominantly in $e^+e^-$, $\mu^+\mu^-$, $\tau^+\tau^-$ and $\bar{b}b$. In this analysis, we will use data from Fermi-LAT, H.E.S.S., Planck and CTA experiments. And, for example, in the Higgs-like case we exclude the canonical annihilation cross section $\langle \sigma v \rangle = 3 \times 10^{-26}$ for a DM mass $m_{\chi} \lesssim 105$~GeV and for $\phi \to e^+e^-$ channel, we have a gain in sensitivity from the CTA experiment of about one order of magnitude, for example, fixing $m_{DM}=5000$~GeV, we have $\langle \sigma v \rangle \sim 2.5 \times 10^{-24}$ for the current limits, while the expected from CTA gives $\langle \sigma v \rangle \sim 1.7 \times 10^{-25}$.
Secluded dark matter is one of the most popular dark matter models, where dark matter annihilations go into particles that belong to a dark sector. An interesting way to probe such models is via indirect detection. In particular, gamma-ray observations are rather promising. Taking into account $1\%$ level of systematics, in this work we show that the Cherenkov Telescope Array (CTA) will place the most stringent bounds on the dark matter annihilation cross section, surpassing existing probes based on H.E.S.S., Fermi-LAT, and Planck data, for dark matter masses above $400$~GeV, independently of the channel and choosing the Einasto profile. We consider scenarios of secluded annihilations into leptophilic, leptophobic scalars and Higgs-like particles being able to exclude an annihilation cross section of $1.7 \times 10^{-25}$~cm$^3 \,$s$^{-1}$ for $m_{DM}=5000$~GeV, for $\phi \to e^+e^-$ channel, for example, exceeding the current limits by one order of magnitude. 
\end{abstract}

%\begin{document}
\maketitle
%\flushbottom

\section{Introduction}
\label{sec:intro}

Weakly Interacting Massive Particles (WIMPs), although well motivated by thermal relic abundance in the electroweak scale, have been strictly constrained by direct and collider searches \cite{Arcadi:2017kky}. For example, XENON1T reaches a scattering cross section of about $4 \times 10^{-47}$~cm$^2$ for a Dark Matter (DM) mass of $30$~GeV \cite{Aprile:2018dbl}, making more and more difficult to find WIMP models able to evade these bounds.

In this way, alternative scenarios have been extensively studied in the literature, between them, new mechanisms for DM production, like freeze-in \cite{Bernal:2017kxu,Bernal:2018hjm,DEramo:2017ecx,Belanger:2018sti}, models with light DM \cite{Frere:2006hp,Hooper:2007tu,Huang:2013zga,Arhrib:2015dez,Boudaud:2016mos,Darme:2017glc,Choudhury:2017osc,Bertuzzo:2017lwt,Dutra:2018gmv}, semi-annihilation models \cite{DEramo:2010keq,Belanger:2012vp,DEramo:2012fou,Aoki:2014cja,Arcadi:2017vis}, secluded models \cite{Pospelov:2007mp}, and so on. In this work, we are going to study these secluded scenarios, where the DM particles couple directly to non-standard mediators, which leads to a suppression in the scattering cross section becoming possible to escape from the current limits, and, at the same time, making the indirect searches important to probe them.

In the face of this, indirect searches appear as a very important way to probe some of these alternative models. Basically, in high-density regions we expect a DM annihilation, and their products, like photons, electrons, neutrinos and antimatter, can propagate from their sources and reach our satellites and telescopes. For example, in secluded models, the DM annihilates in meta-stable mediators that subsequently decay into standard model particles. Here, we will focus on the gamma-ray products of these annihilations, since it has the advantage that, as a neutral particle, it does not deviate during propagation through the galaxy, pointing directly to their sources \cite{Bringmann:2012ez}.

Secluded scenarios appear in different contexts in the literature, for example, explaining the gamma-ray excess observed in the Galactic Center (GC), in some cases choosing the mediator as a gauge boson from an extra $U'(1)$ symmetry \cite{Fortes:2015qka}, as well as when the mediator is very light, in the called eXciting DM (XDM) \cite{Finkbeiner:2007kk,ArkaniHamed:2008qn,Calore:2014nla}, and via scalar fields \cite{Kopp:2013mi,Kim:2016csm,Elor:2015tva}.
Solar gamma-ray and neutrino searches with experiments like HAWC \cite{Albert:2018jwh}, Icecube \cite{Aartsen:2016zhm}, Antares \cite{Adrian-Martinez:2016gti} and super-K \cite{Choi:2015ara} provide another way to look for secluded scenarios \cite{Batell:2009zp,Allahverdi:2016fvl,Leane:2017vag}, since only neutrinos can escape from the Sun's atmosphere in direct DM annihilation \cite{Aartsen:2016zhm}. Another important work in the context of indirect searches including hidden sectors, chosen n-cascade annihilation via scalars (and vectors\footnote{Reference \cite{Elor:2015tva} showed that the change in the spectrum for scalars and vectors mediators is almost insignificant.}), used data from Planck, via Cosmic Microwave Background (CMB), from Fermi-LAT using gamma-ray data from dwarfs, and positron data from AMS-02 in order to constrain these secluded scenarios \cite{Elor:2015tva}. In addition, complementary searches were made via fermion as the mediator \cite{Campos:2017odj}, and scalar or vector mediators \cite{secluded}. Summarizing, secluded scenarios can appear in different models, including minimal extensions of the Standard Model (SM) as well as supersymmetric models \cite{Martin:1997ns,Nomura:2008ru,Araz:2018uyi} and can be probed in a number of ways.

In this paper, we are doing for the first time the combined analysis for secluded models where the mediator is a Higgs-like particle in a model-independent way. We use data from Fermi-LAT looking at dwarf spheroidal galaxies (dSphs) \cite{Ackermann:2015zua}, from H.E.S.S. data searching for gamma-rays from the Galactic Centre (GC) \cite{Abdallah:2016ygi}, and Planck data for the cosmic microwave background (CMB) \cite{Aghanim:2018eyx}, to compare with the prospects for CTA looking at the GC direction \cite{Silverwood:2014yza}. In addition, we complement the analysis from \cite{secluded}, including for the first time the CTA prospects for the mediators decaying predominantly into $e^+e^-$, $\mu^+\mu^-$, $\tau^+\tau^-$ or $\bar{b}b$, in order to check what will be the enhancement in sensitivity in the next generation of high energy gamma-rays by CTA, which is the main focus of this work.\\

The paper will be structured in the following way: in the Section \ref{sec:dmsecluded}, we briefly introduce the secluded models, including the description of the scenarios approached in this work; in Section \ref{sec:indirect}, we describe the indirect detection techniques, including the computation of the gamma-ray spectrum and flux; in Section \ref{sec:expdata}, we discuss the existing limits; in Section \ref{sec:results}, we present the CTA experiment and the results and, in the last Section, we present our conclusions.
 
\section{Secluded Dark Matter}
\label{sec:dmsecluded}

\begin{figure}[ht]
    \centering
    \includegraphics[scale=0.3]{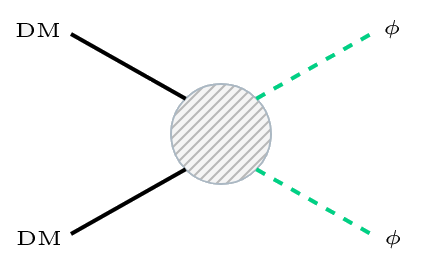}
    \caption{Feynman diagram for secluded models with scalar mediators.}
    \label{fig:diag}
\end{figure}

%Secluded Models are gain visibility due to the lack of results from collider and direct searches for WIMPs and it has been overly studied in different contexts in the literature
Secluded dark matter models appear as a good alternative to evade the strong current limits from direct and collider searches, and it has been overly studied in different contexts in the literature
\cite{Pospelov:2007mp,Essig:2009jx,Martin:2014sxa,Alves:2015pea,Ducu:2015fda,Alves:2015mua,Okada:2016tci,Duerr:2016tmh,Jacques:2016dqz,Karam:2015jta,Karam:2016rsz,Arcadi:2016kmk,Cirelli:2016rnw,Baldes:2017gzu,Arcadi:2017jqd,Balducci:2017vwg,Evans:2017kti,secluded,Breitbach:2018ddu,Balducci:2018dms,Balducci:2018ryj,Cirelli:2018iax}. In this kind of scenario, we can easily avoid these constraints, and indirect detection searches become important to probe these models, this is because the DM does not couple directly to Standard Model particles as usually, instead, the DM particles couple to `dark mediators' which can subsequently decay in SM particles. Among several possibilities that provide this setup, the most simple one is imposing the DM mass to be larger than the mass of the mediator, and if the DM particle has a strong enough coupling to the mediators, the channel in Fig.~\ref{fig:diag} becomes predominant leaving the DM secluded \cite{Pospelov:2007mp}, another one is via Chern-Simons Portal or via kinetic mixing as discussed in \cite{Arcadi:2017jqd}, between others.
%
%the conditions that lead to Secluded scenarios are the relation between the masses of the DM particle and the mediator and the coupling between them is strong enough, i.e., when the DM mass is larger than the mass of the mediator, the channel in Fig.~\ref{fig:diag} becomes predominant, consequently, the coupling between SM and DM particles, indirect.

There are several possibilities for this mediator, it can be a scalar, a vector or a fermion, as discussed in \cite{Pospelov:2007mp}. In this work, we choose the mediator as scalar particles for several scenarios, among them as a Higgs-like scalar, or a leptophilic scalar, where the scalar decays predominantly into $e^+e^-$, $\mu^+\mu^-$ or $\tau^+\tau^-$, or, a leptophobic scalar, which decays mainly into $\bar{b}b$. We will discuss, for each scenario, how they are constrained by the current and future experiments in indirect DM searches.

\section{Indirect searches} 
\label{sec:indirect}

\subsection{Gamma-ray Spectrum}

One of the most important ingredients that distinguish different DM particle models from the background in indirect searches is the gamma-ray spectrum, $dN^\gamma/dE$, produced in DM annihilations. It depends on the DM particle mass and of the primary channel. Depending on the model, this quantity will be given by the sum over the branching fraction for each channel,
\begin{equation}
    \frac{dN^\gamma}{dE}= \sum_{i} BR_i \times \frac{dN^\gamma_i}{dE},
    \label{dnde}
\end{equation}
where $i$ runs over all the channels involved in the process. For example, the Higgs-like particle with $100$~GeV, considered in this project, has a branching ratio very close to the standard case, while when the scalar mass is $10$~GeV the predominant decay is about $\sim 70\%$ in $\sim \bar{b}b$ and $~20\%$ in $\tau^+\tau^-$, and this will determine the shape of the spectrum and consequently the possible observed flux in our detectors, as we will see in details in the next Section. In the following, we will present our results for the gamma-ray spectrum for the different scenarios.  

\subsubsection{Mediator $\phi + \phi$ (Higgs-like)}

The first scenario considered here is the case where the mediator is a Higgs-like scalar $\phi$, i.e., a scalar which couples to the SM particles like the standard Higgs, just varying the mediator's mass, $m_\phi=10$~GeV, $m_\phi=100$~GeV, and $m_\phi=500$~GeV. For each scalar mass, we computed the branching ratios using SARAH and SPheno packages \cite{Staub:2011dp,Staub:2013tta,Staub:2015kfa,Porod:2003um,Porod:2011nf}.

\begin{figure}[ht]
\includegraphics[width=0.95\columnwidth]{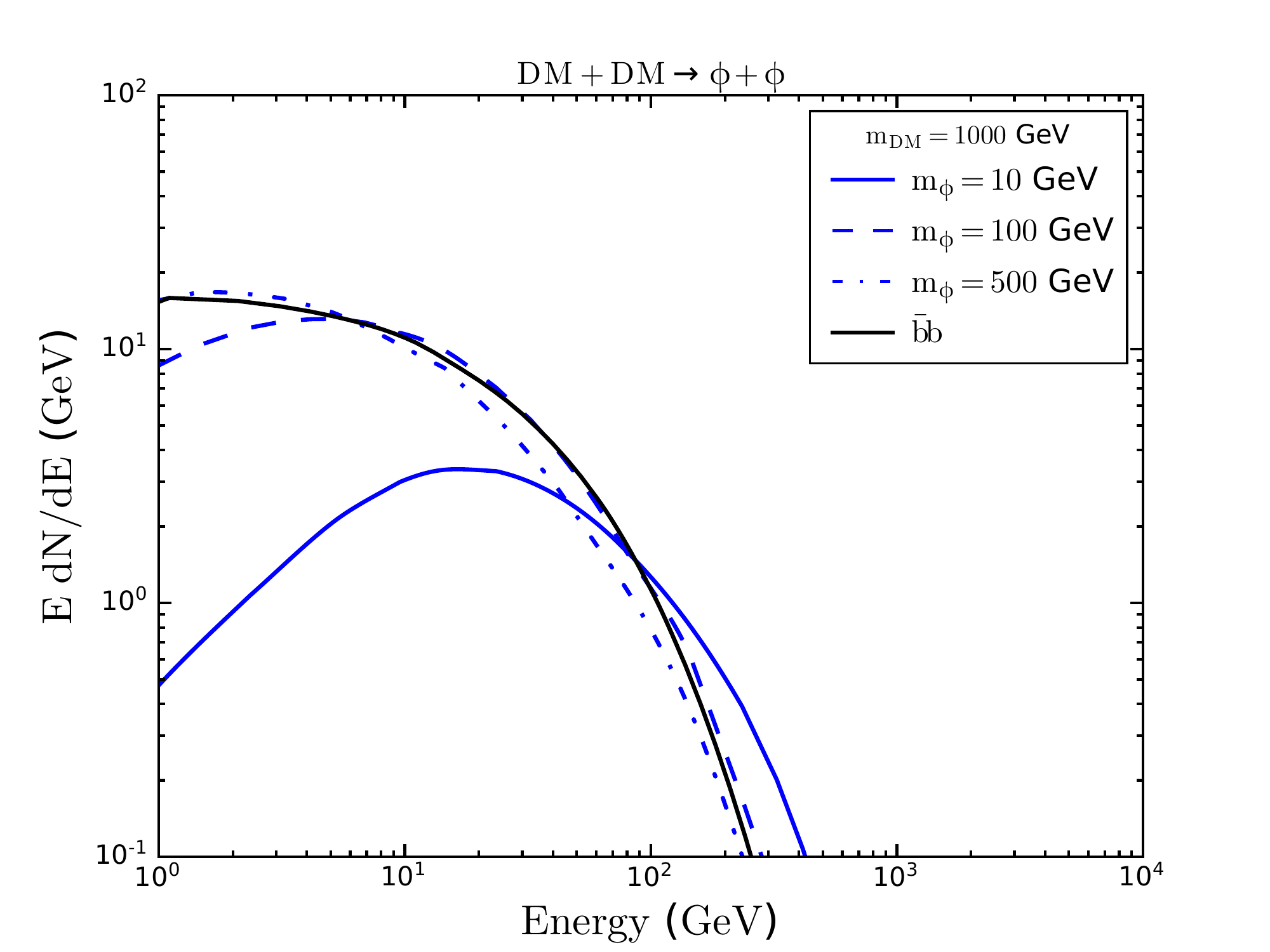}
\caption{$E\times dN/dE$ \textit{versus} energy for the $\phi$ as a Higgs-like particle, the DM mass was taken equal to $1000$~GeV, and the mediator mass equal to $10$~GeV (continuous lines), $100$~GeV (dashed lines) and $500$~GeV (dashed-dotted lines).}
\label{dndephihiggslike}
\end{figure}

In the Fig.~\ref{dndephihiggslike}, we present the photon spectrum for DM particles annihilating in Higgs-like particles $DM + DM \rightarrow \phi + \phi$, computed numerically by Pythia 8 package \cite{Sjostrand:2014zea}. We impose the DM mass equal to $1000$~GeV and, we take the scalar mass equal to $10$~GeV (continuous lines), $100$~GeV (dashed lines) and $500$~GeV (dashed-dotted lines). As discussed before, our $100$~GeV mass scalar is very close to the standard one, and this includes a significant branching ratio into $\gamma\gamma$ and into $\bar{b}b$. In the others, for masses equal to $10$~GeV and $500$~GeV, the branching ratio into $\gamma\gamma$ becomes suppressed, and the most important channels are $\bar{b}b$ ($\sim 70\%$) and $\tau^+\tau^-$ ($\sim 20\%$), for $m_\phi = 10$~GeV, and  $W^+W^-$ ($\sim 55\%$), $\bar{t}t$ ($\sim 20\%$) and $ZZ$ ($\sim 15\%$) for $m_\phi = 500$~GeV. The first case ($m_\phi = 10$~GeV), provides a spectrum mainly dominated by pions, since both $\bar{b}b$ and $\tau^+\tau^-$ quickly hadronize producing pions which decay instantaneously into $\gamma\gamma$, the spectrum is harder compared to the direct production into $\bar{b}b$ (black line) due to $\phi$ be produced boosted, namely, more high energetic photons are produced during the annihilation. For higher scalar masses, the spectrum follows the same behavior but it becomes softer compared to the smallest case.

\begin{figure}[ht]
\centering
\includegraphics[width=0.95\columnwidth]{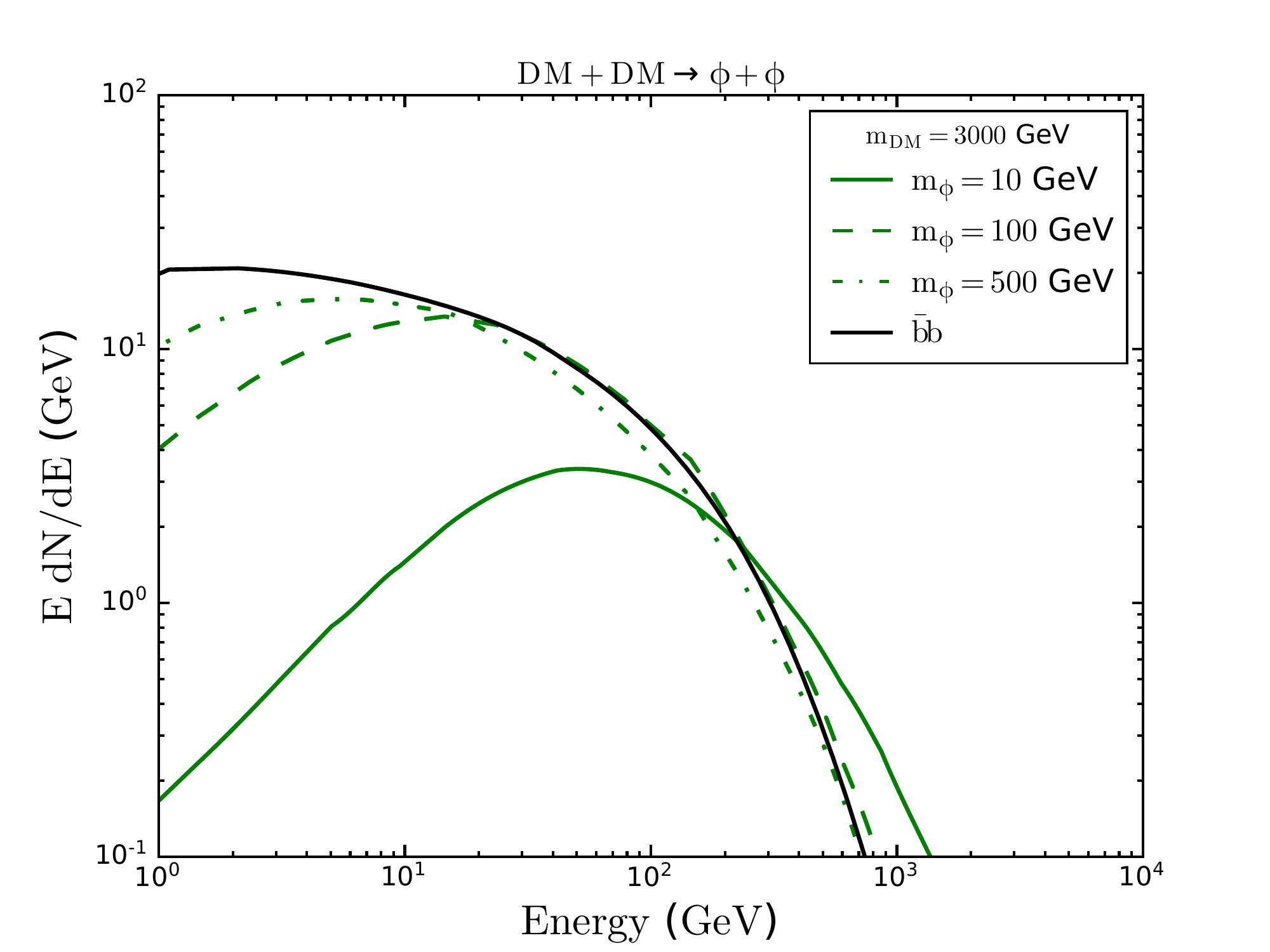}
\caption{$E\times dN/dE$ \textit{versus} energy for $\phi$ as a Higgs-like particle, the DM mass was taken equal to $3000$~GeV, and the mediator mass equal to $10$~GeV (continuous lines), $100$~GeV (dashed lines) and $500$~GeV (dashed-dotted lines), as high as the DM mass is, more energetic gamma-rays will be produced. For comparison, we include the spectrum directly into $\bar{b}b$ (black line).}
\label{dndephihiggslike3000}
\end{figure}

In the Fig.~\ref{dndephihiggslike3000}, we show the same results just changing the mass of the DM particle to $3000$~GeV, in this case, it is evident that the shape of the spectra is the same but it is moved up, showing that more energetic photons are produced, given a harder spectrum. The same analyses mentioned before applies here.

\subsubsection{Mediator $\phi + \phi$ decaying predominantly in $e^+e^-$, $\mu^+\mu^-$, $\tau^+\tau^-$ or $\bar{b}b$}

%In this case, in order to be more general as possible, we consider the mediator decaying predominantly in $e^+e^-$, $\mu^+\mu^-$, $\tau^+\tau^-$, which we call leptophilic scalar, or $\bar{b}b$, the leptophobic scalar.
In this section we considered the leptophilic scalar, which decays exclusively in $e^+e^-$, $\mu^+\mu^-$ or $\tau^+\tau^-$, and the leptophobic scalar, which decays predominantly in $\bar{b}b$.
As discussed previously, each one generates a specific spectrum leading the signature of the model. Using this spectrum, we can compute the expected DM fluxes, enabling us to get the upper limits on the annihilation cross section for each model.
\begin{figure}[ht]
\centering
\includegraphics[width=0.95\columnwidth]{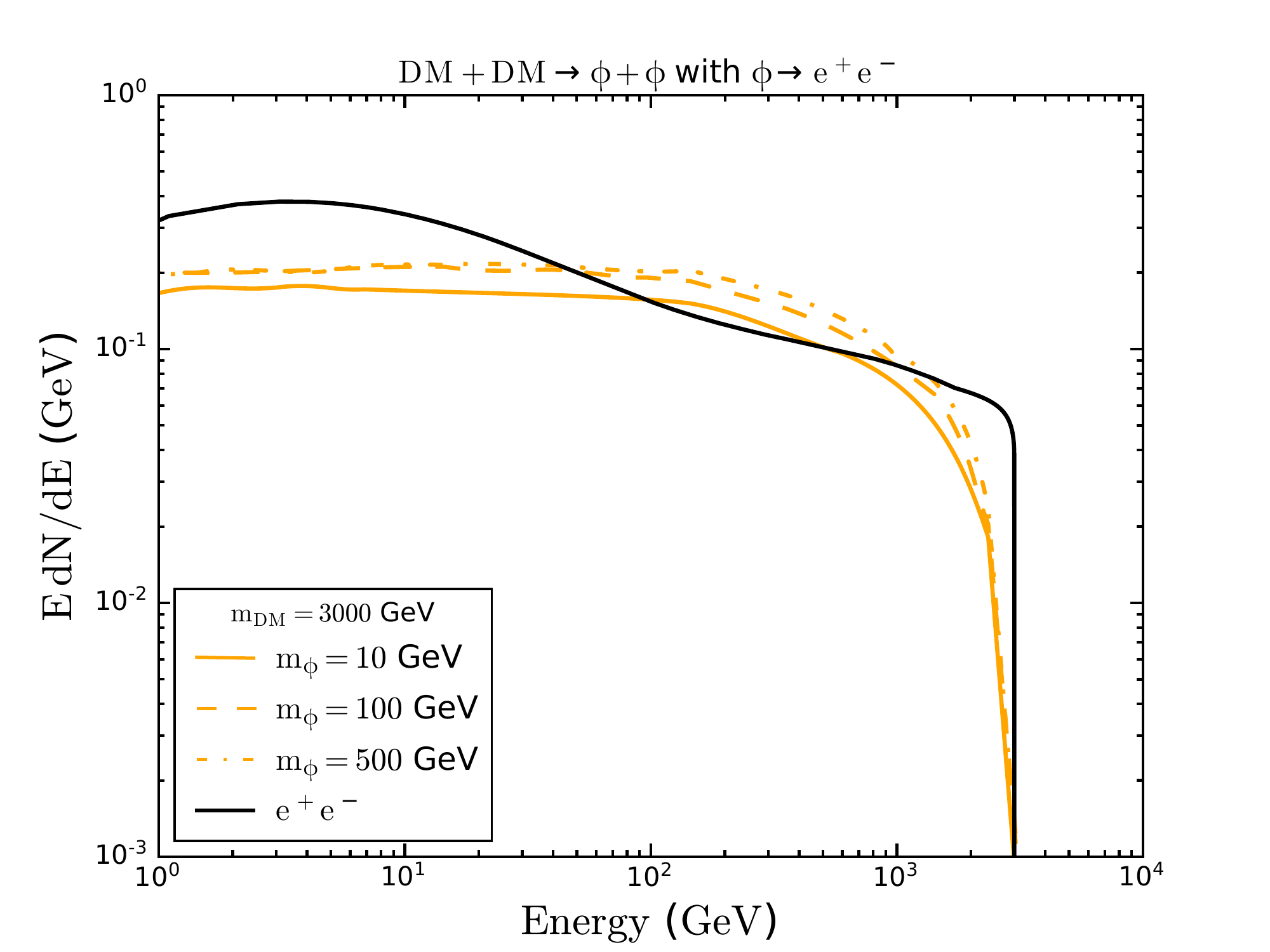}
\caption{$E\times dN/dE$ \textit{versus} energy for the mediator $\phi$ decaying predominantly in $e^+ e^-$, the DM mass is equal to $3000$~GeV, and assuming following values for the mediator mass: $10$~GeV (continuous lines), $100$~GeV (dashed lines) and $500$~GeV (dashed-dotted lines). For comparison, we include the spectrum directly into $e^+e^-$ (black line).}
\label{dndephiphiee}
\end{figure}

\begin{figure}[ht]
\centering
\includegraphics[width=0.95\columnwidth]{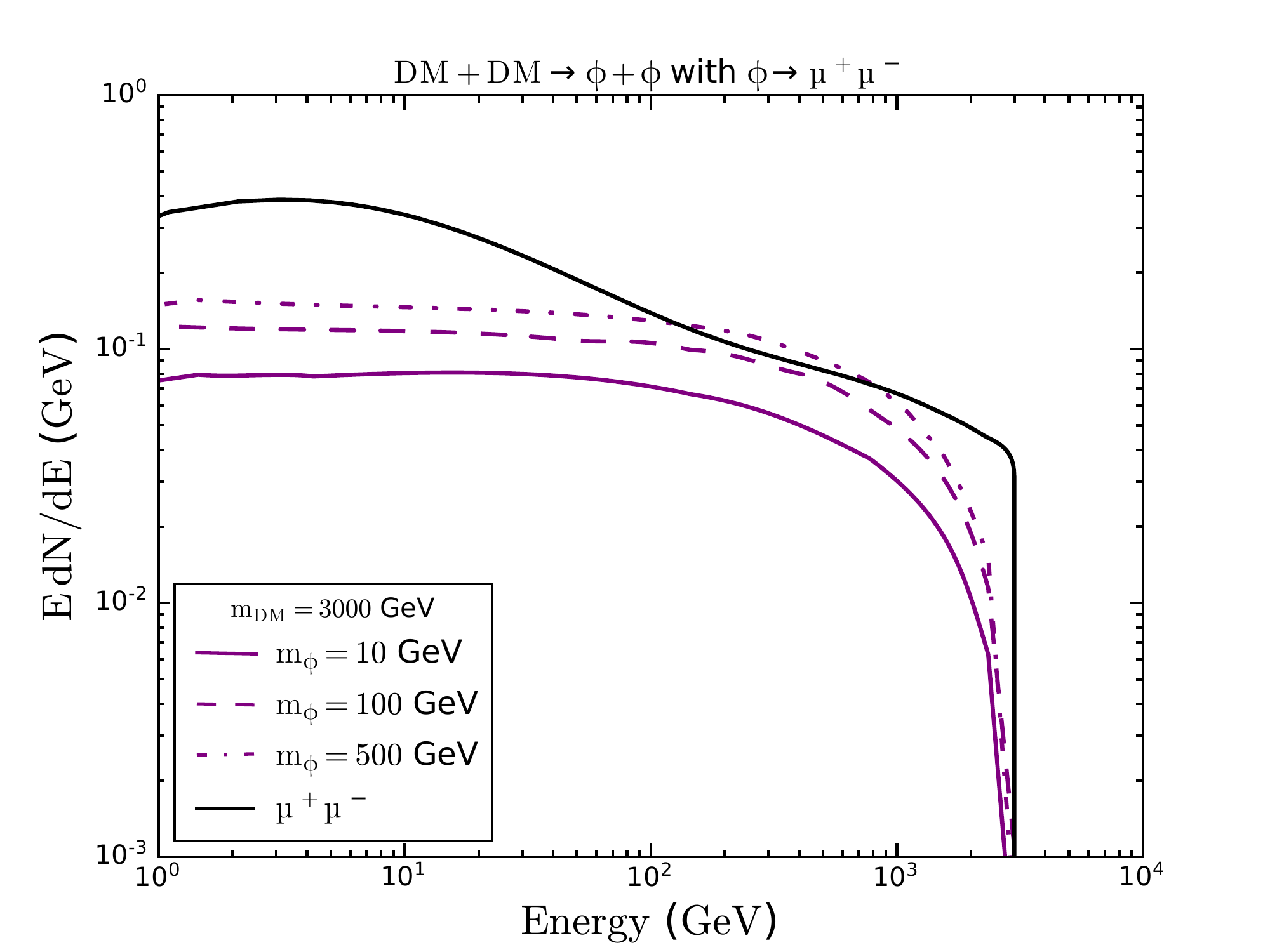}
\caption{$E\times dN/dE$ \textit{versus} energy for the mediator $\phi$ decaying predominantly into $\mu^+ \mu^-$, here DM mass is $3000$~GeV, and assuming following values for the mediator mass: $10$~GeV (continuous lines), $100$~GeV (dashed lines) and $500$~GeV (dashed-dotted lines). For comparison, we include the spectrum directly into $\mu^+\mu^-$ (black line).}
\label{dndephiphimumu}
\end{figure}

\begin{figure}[ht]
\centering
\includegraphics[width=0.95\columnwidth]{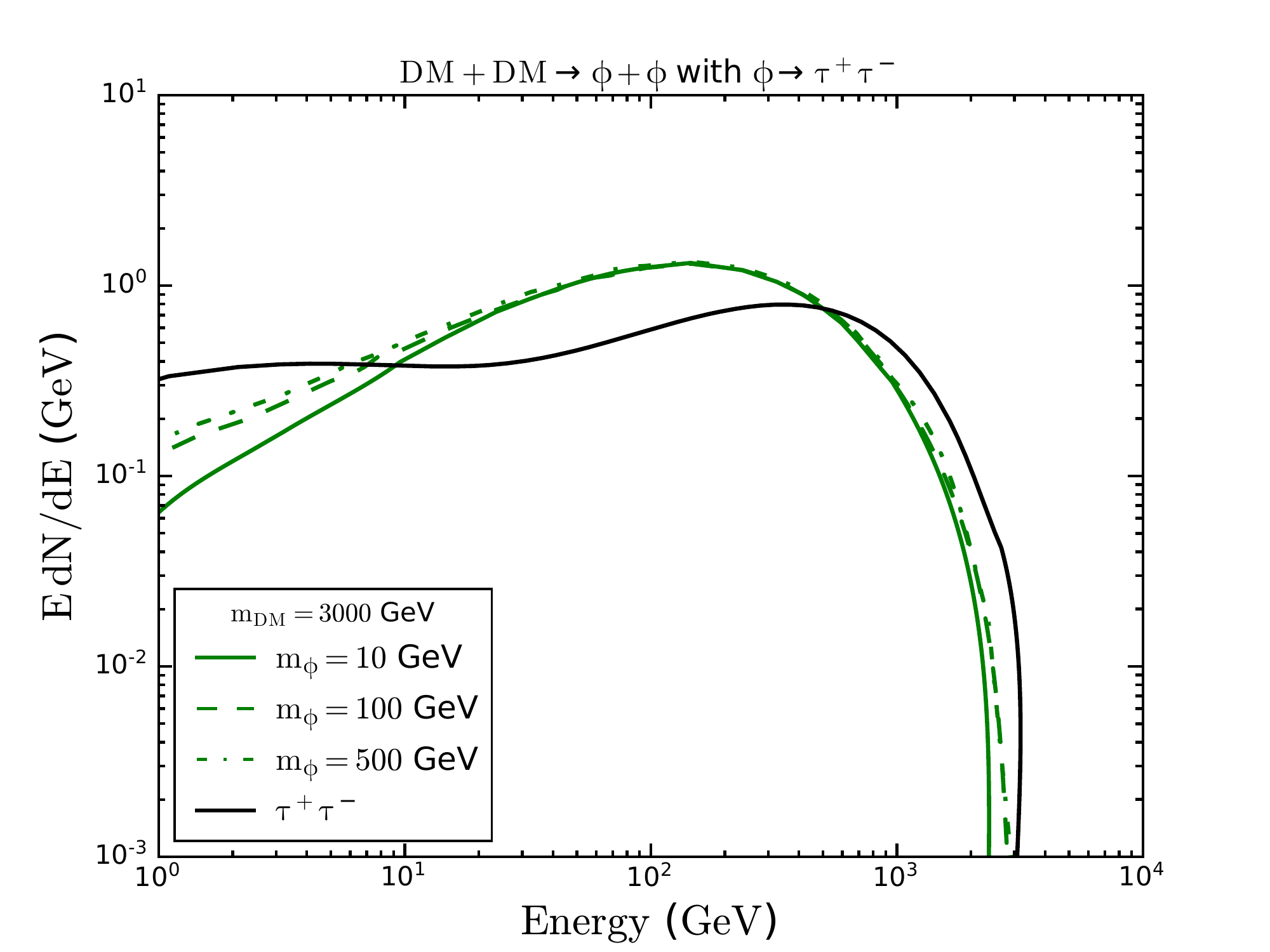}
\caption{$E\times dN/dE$ \textit{versus} energy for the mediator $\phi$ decaying predominantly in $\tau^+ \tau^-$, for the DM mass equal to $3000$~GeV, and assuming following values for the mediator mass: $10$~GeV (continuous lines), $100$~GeV (dashed lines) and $500$~GeV (dashed-dotted lines). For comparison, we include the spectrum directly into $\tau^+\tau^-$ (black line).}
\label{dndephiphitautau}
\end{figure}

\begin{figure}[ht]
\centering
\includegraphics[width=0.95\columnwidth]{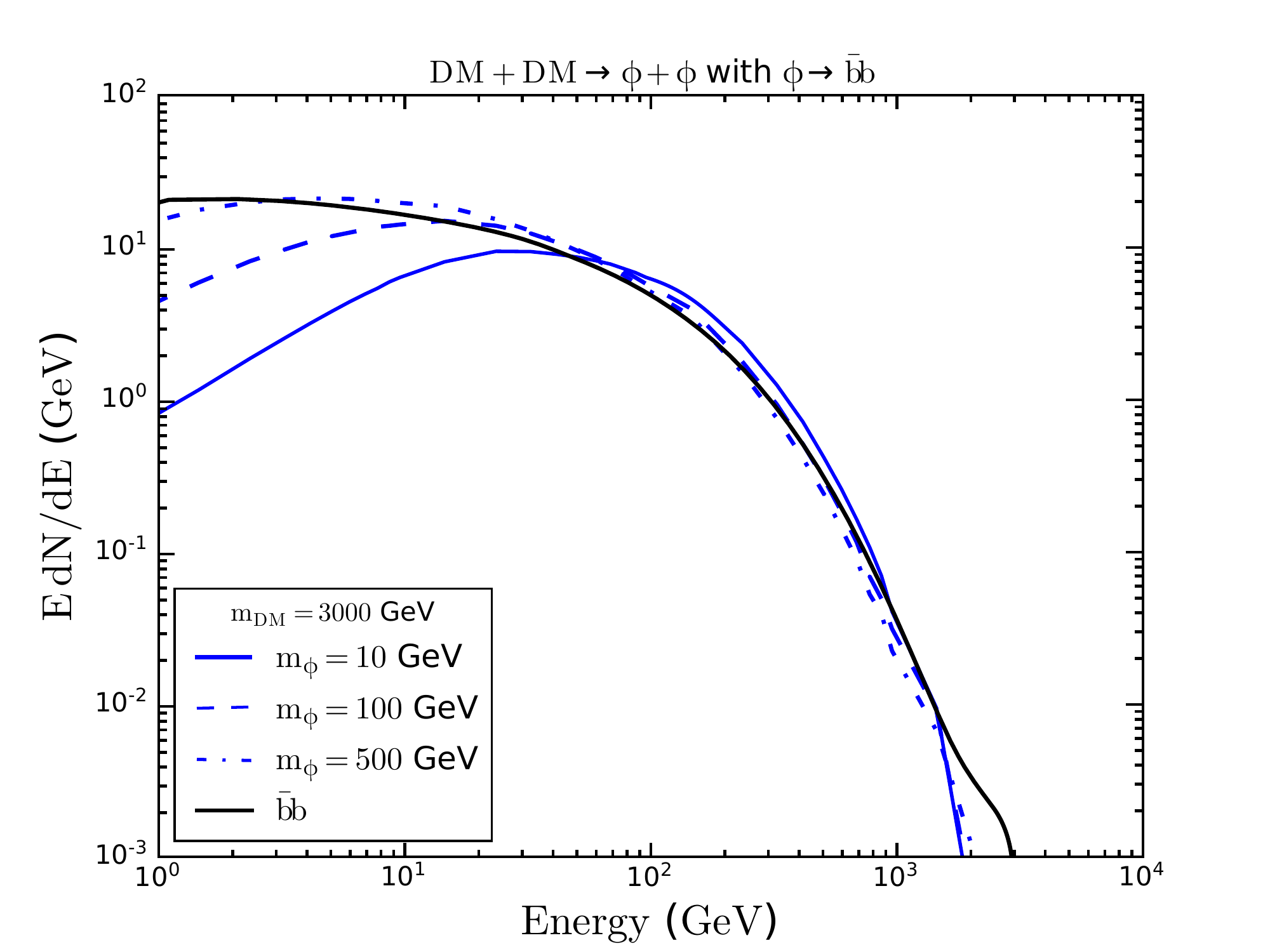}
\caption{$E\times dN/dE$ \textit{versus} energy for the mediator $\phi$ decaying predominantly in $\bar{b}b$, for DM mass equal to $3000$~GeV, and assuming following values for the mediator mass: $10$~GeV (continuous lines), $100$~GeV (dashed lines) and $500$~GeV (dashed-dotted lines). For comparison, we include the spectrum directly into $\bar{b}b$ (black line).}
\label{dndephiphibb}
\end{figure}

The production of gamma-rays via leptonic channels, like $e^+e^-$ and $\mu^+\mu^-$ are predominantly given by final state radiation (FSR), which provides a hard spectrum with a sharp edge in the DM mass, in the other side, for the final states $\tau^+\tau^-$ and $\bar{b}b$, the process of hadronization and fragmentation producing pions which decays almost $100\%$ in photons, given a major number of photons but softer compared to the leptonic decays \cite{Birkedal:2005ep,Slatyer:2017sev}. These spectral shapes can be seen in Figs.\ref{dndephiphiee}-\ref{dndephiphibb}, where we provide for each channel a comparison with the standard case without a dark mediator.

In Fig.~\ref{dndephiphiee}, we show the spectrum for the case where the mediator decays predominantly in $e^+e^-$, we take the DM mass equal to $3000$~GeV and vary the mass of the mediator equal to $10$~GeV, $100$~GeV and $500$~GeV, obviously, the shape of the spectrum is the same, but with a slight increase in the $dN^\gamma/dE$ due to the variation in the mediator $\phi$ mass.

In the second case, we choose the mediator decaying into $\mu^+\mu^-$ (see Fig.~\ref{dndephiphimumu}), and we follow the same values for the masses as before, the shape of the spectrum is very similar to the last case, basically following the characteristic leptonic decay. Varying the mass of the mediator leads to a bigger split in the spectrum than in the electronic one.

In the third case, we take the mediator decaying in $\tau^+\tau^-$ (see Fig.~\ref{dndephiphitautau}), here we observe a considerable difference in the shape of the spectrum relative to the other leptonic channels, this is because $\tau$ decays predominantly in hadrons, as explained before. In addition, the effect of the change in the mediator's mass is mild.

In the last case, we take the $\bar{b}b$ as a final state (see Fig.~\ref{dndephiphibb}), assuming the same values for the masses as before. Here we observe a major difference by varying the mediator mass, especially for lower energies. It is important to emphasize that the shape of the spectrum will give the signature of the channel in question. In addition, all the gamma-ray spectra were computed using the numerical package Pythia 8 \cite{Sjostrand:2014zea}.

Now, with the main ingredients in hands, we can follow to calculate the gamma-ray fluxes for DM annihilation.

\subsection{Gamma-ray flux}
\label{sec:fermi}

Gamma-rays are very interesting signals in the DM search, mainly because they trace back the sources, in the case of charged particles this information is usually lost due to the propagation, moreover it is easier to detect compared to neutrinos. Therefore, gamma-rays will be the main aim in indirect searches addressed here.

The gamma-ray flux expected from a DM annihilation is given by,
\begin{equation}
 \frac{d\phi_\gamma}{dE}=\frac{r_\odot}{8\pi}\left(\frac{\rho_\odot}{m_{DM}}\right)^2 \langle \sigma v \rangle  \frac{dN_\gamma}{dE} J_{Ein}
\end{equation}
where $m_{DM}$ is the DM mass, $ \langle \sigma v \rangle $ is the velocity averaged annihilation cross section times velocity, $dN_\gamma/dE$ is the gamma-ray spectrum, showed in Eq.~(\ref{dnde}) as a sum over the primary final states, and $J_{Ein}$ is the J-factor for annihilation choosing the Einasto halo profile \cite{Graham:2005xx,Navarro:2008kc}, 
\begin{equation}
 J_{Ein}=\int_{l.o.s}\frac{ds}{r_\odot}\left(\frac{\rho_{Ein}(r(s,\theta))}{\rho_\odot}\right)^2,
\end{equation}
where $r_\odot=8.33$~kpc and $\rho_\odot=0.4$~GeV$/$cm$^3$ are the distance between the sun and the Galactic Centre (GC) and the DM density in the location of the Sun, respectively. And, the DM density for the Einasto profile \cite{Graham:2005xx,Navarro:2008kc}, 
\begin{equation}
    \rho_{Ein}(r) = \rho_{s} \exp{ \left( \frac{-2}{\alpha}\left[\left(\frac{r}{r_{s}}\right)^{\alpha}-1 \right] \right)}
\end{equation}
with $r$ being the galactic radius, and $r_s=28.44$~kpc, $\rho_s=0.033$~GeV$/$cm$^3$ and, $\alpha=0.17$, the typical constants for Einasto profile (for more details, please, see \cite{Cirelli:2010xx}). 

Afterward, we are going to use data from Fermi-LAT \cite{Ackermann:2015zua}, H.E.S.S. \cite{Abdallah:2016ygi} and Planck \cite{Ade:2015xua}, in order to verify the impact of prospects for the CTA experiment \cite{Silverwood:2014yza} on the survey of the aforementioned secluded models. To start, we will describe the existing limits, the CTA datasets, and the methodology to get our bounds.

\section{Existing limits}
\label{sec:expdata}

\subsection{Fermi-LAT}

During the 6 years of dSphs' observations by Fermi-LAT experiment no gamma-ray excess has been observed so far, in the energy range between $500$~MeV to $500$~GeV, putting strong limits over the DM annihilation cross section for the standard channels, like $\bar{b}b$ and $\tau^+ \tau^-$, using the PASS 8 event-level analysis \cite{Ackermann:2015zua}. In particular, in this paper the collaboration report for the $\bar{b}b$ channel, the exclusion of the canonical annihilation cross section, $3 \times 10^{-26}$~cm$^3 \,$s$^{-1}$, for DM masses below $100$~GeV.

In this work, we will follow the same recipe described in \cite{Ackermann:2015zua}, and in order to compute the limits for the Higgs-like case, we will use the gamLike routine from the GAMBIT package \cite{Workgroup:2017lvb}, to compare and check the impact of the CTA sensitivity on these models. 

\subsection{\textit{High Energy Stereoscopic System} (H.E.S.S.)}

The H.E.S.S. experiment is a ground-based telescope located in Namibia, which is looking for high energy gamma-rays at the Galactic Centre. Able to explore an energy interval between $230$~GeV until $30$~TeV, H.E.S.S. gives, currently, the most stringent bounds in which concerns DM searches with high energy gamma-rays. As an example, using data from $254$h of exposure time, strong limits were imposed over the DM annihilation cross section for the standard channels~\cite{Abdallah:2016ygi}, specifically, for the $\tau^+ \tau^-$ channel a limit of about $\langle \sigma v \rangle \sim 1 \times 10^{-26}$~cm$^3 \,$s$^{-1}$ for a DM mass of around $1000$~GeV was placed.

The last results from H.E.S.S. does not provide the necessary information to reproduce their results, however, following the receipt described in \cite{Profumo:2016idl}, implemented in the package gamLike, it is possible to reproduce their limits for $112$~h observation time with very good agreement \cite{Workgroup:2017lvb}, where the methodology consists on the separation of two regions of the sky called \textit{background} and \textit{signal} (expected to observe DM signals) regions, excluding the $|b|<0.3^\circ$ in order to avoid the high intensity of gamma-rays region. Through these limits, we were able to reproduce the results for $254$~h, by a simple re-scaling of the $112$~h limits, which gives a very good approximation as showed in \cite{secluded}.

In order to compute the limits for the Higgs-like case, we follow the description above, where we compute the likelihood function in order to get the exclusion curve, based on $112$h exposure~\cite{Abramowski:2011hc}. Afterward, due to the lack the necessary information in order to recover their $254$h results, we just re-scale our limits for $112$h (in a good approximation, we reproduced their channels) in order to get an approximated exclusion line based on $254$h exposure, and this will be the result presented in the Section~\ref{sec:results} to compare with the CTA limits. It is important to emphasize that this analysis, although reproduces very well the standard limits from the collaboration, gives just an approximation.  

\subsection{Planck Satellite}

The Planck satellite gave us another view on the CMB radiation, given precise measurements of its anisotropies, then everything that affects the CMB is strongly constrained by the last results of the Planck satellite. Therefore, the constraint over the annihilation parameter is given by \cite{Aghanim:2018eyx},
\begin{equation}
p_{ann} < 3.2 \times 10^{-28}  \, \textrm{cm}^3 \, \textrm{s}^{-1} \, \textrm{GeV}^{-1} 
\end{equation}
and using this value in,
\begin{equation}
p_{ann}=f_{eff}\frac{\langle \sigma v \rangle}{m_{DM}},
\label{pann}
\end{equation}
we compute the limit over the DM annihilation cross section versus DM mass, calculating the efficiency function for secluded models, that provides the relation between the energy injected and deposited in the thermal bath, following the prescription and using the routines provided in \cite{Slatyer:2015jla}.\\

Using these limits we will be able to verify the impact of the CTA experiment over the secluded models. We will describe the details of the CTA experiment and present our results in the next Section.

\section{The Cherenkov Telescope Array (CTA)}
\label{sec:results}

The upcoming experiment in high energy gamma-rays, the Cherenkov Telescope Array (CTA), will cover an energy range between $\sim 20$~GeV to $\sim 300$~TeV. The experiment will be composed of three types of telescopes, with Small, Medium and Large sizes, to be able to explore from high to low energies, respectively. In addition, it will be located in the North (La Palma) and South (Chile) hemispheres in order to have a wider view of the sky, but the South will give a better view of DM in the Galactic Centre.

In this work, we compute the expected sensitivity of the CTA looking in the direction of the Galactic Centre to the secluded models, following the morphological analysis, as described in \cite{Silverwood:2014yza}, where they separate the Region of Interest (RoI) in 28 sub-RoIs. In this analysis, they are taking into account a configuration of telescopes with 3 Larger, 18 Medium and 56 Small telescopes, given an effective area of about $100$~m$^2$, with $100$h observation time, as well as, taking into account cosmic-ray and galactic diffuse emission (GDE) as possible backgrounds.

Instead of follow other analysis including that made by the collaboration \cite{Doro:2012xx,Pierre:2014tra,Morselli:2017ree} which gives an optimistic scenario, we choose this methodology described above in order to be conservative, previous analysis do not consider GDE and/or do not take into account systematic uncertainties (related, for example, to the acceptance of the experiment in a Field of View (FoV)), here we are considering both possibilities, therefore, we are following a conservative approach as given by \cite{Silverwood:2014yza}.

Following the previous description and using the gamLike package \cite{Workgroup:2017lvb}, we compute the binned Poisson likelihood function, then we use the statistical test $-2\Delta \ln{\mathcal{L}<2.71}$ in order to get the $95\%$~C.L. exclusion limits, with $1\%$ level of systematics.\\

Using the techniques described above, we are able to compute the upper limits over the DM annihilation cross section \textit{versus} DM mass for secluded models in the following setups, in the first one, choosing the mediator as a Higgs-like scalar with masses $10$~GeV, $100$~GeV and $500$~GeV, i.e., coupling to the standard particles as in the standard case. In the second case, we study the possibility of the mediator as a leptophilic scalar, that couples predominantly to $e^+e^-$, $\mu^+\mu^-$ or $\tau^+\tau^-$, or a leptophobic scalar, that decays into $\bar{b}b$, in the last case, the analysis will be complementary to \cite{secluded}, then it will include the improvement on the sensitivity given by the CTA experiment.

\subsection{Mediator $\phi + \phi$  (Higgs-like)}

In this section, we present our upper limits on the annihilation cross section for the case where the mediator $\phi$ is a Higgs-like particle. First of all, we choose three scenarios changing the mass of the scalar $\phi$, $m_\phi = 10$~GeV, $100$~GeV and $500$~GeV, of course, the branching fractions will change accordingly. In each case, the branching ratio is predominantly composed by hadronic final states. We applied here four different datasets, from Fermi-LAT looking at dSphs (green lines), from H.E.S.S. for the galactic centre (cyan lines), from Planck for CMB searches (black lines) and the prospects for CTA (blue lines).
\begin{figure}[ht]
\centering
\includegraphics[width=0.95\columnwidth]{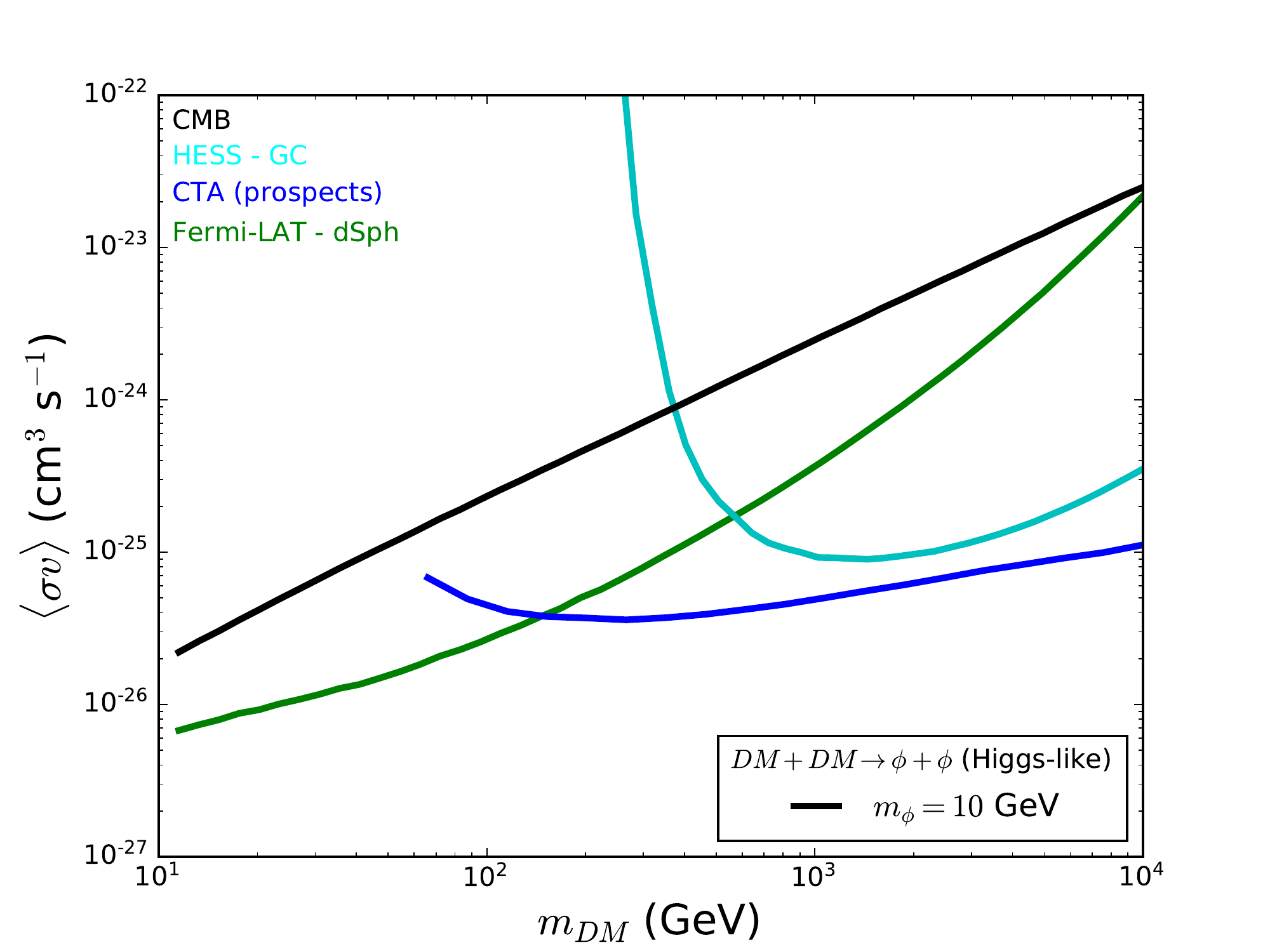}
\caption{Upper limits on the DM annihilation cross section, with $95\%$~C.L., choosing the mediator as a Higgs-like particle with mass equal to $10$~GeV, using the current data from Planck satellite (black lines), from H.E.S.S. experiment for the galactic centre (cyan lines), and from Fermi-LAT experiment for dSphs (green lines), and the CTA prospects (blue lines).}
\label{phihiggslike10}
\end{figure}

\begin{figure}[ht]
\centering
\includegraphics[width=0.95\columnwidth]{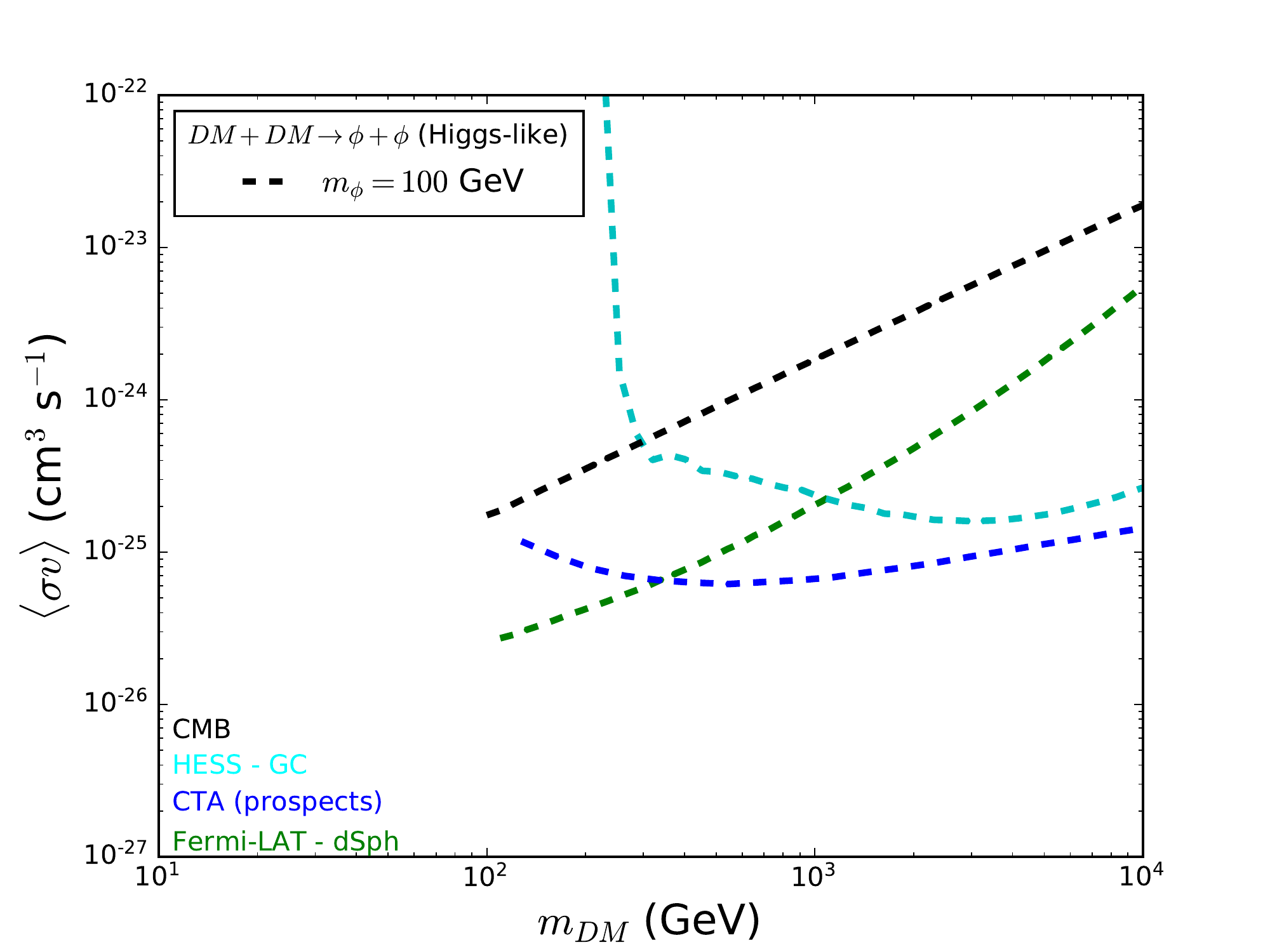}
\caption{Upper limits on the DM annihilation cross section, with $95\%$~C.L., choosing the mediator as a Higgs-like particle with mass equal to $100$~GeV, using the current data from Planck satellite (black lines), from H.E.S.S. experiment for the galactic centre (cyan lines), and from Fermi-LAT experiment for dSphs (green lines), and the CTA prospects (blue lines).}
\label{phihiggslike100}
\end{figure}

Due to the high branching ratio in hadronic final states, the CMB constraints are suppressed relative to the other indirect DM searches. It is important to emphasize that the energy sensitivity of each experiment will be reflected in the limits, for example, the Fermi-LAT is more sensible to lower energies while the H.E.S.S. to high energies, therefore, the lower DM masses will be more constrained by Fermi-LAT and so on.
\begin{figure}[ht]
\centering
\includegraphics[width=0.95\columnwidth]{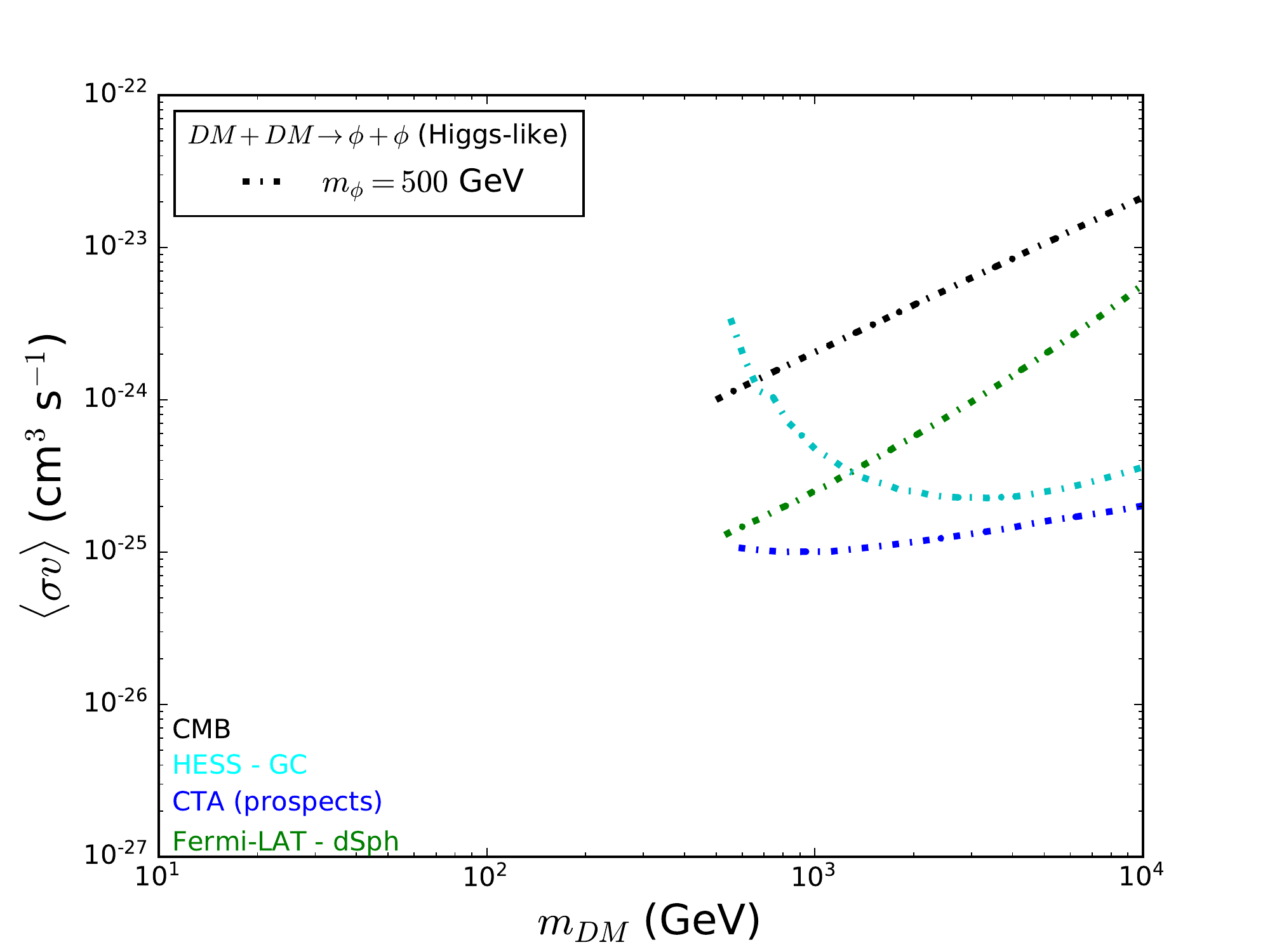}
\caption{Upper limits on the DM annihilation cross section, with $95\%$~C.L., choosing the mediator as a Higgs-like particle with mass equal to $500$~GeV, using the current data from Planck satellite (black lines), from H.E.S.S. experiment for the galactic centre (cyan lines), and from Fermi-LAT experiment for dSphs (green lines), and, using the CTA prospects (blue lines).}
\label{phihiggslike500}
\end{figure}

In the Fig.~\ref{phihiggslike10}, we show our results for $m_\phi=10$~GeV, where the predominant branching fraction is about $70\%$ in $\bar{b}b$ and $20\%$ in $\tau^+\tau^-$, as the lepton $\tau$ decays predominantly in hadrons, the gamma-ray spectrum is mainly hadronic, and we have the CMB bounds poor compared to the other indirect searches. In this analysis, we can exclude the canonical annihilation cross section, $\langle \sigma v \rangle = 3 \times 10^{-26}$~cm$^3 \,$s$^{-1}$, (that provides the correct relic density) for DM masses below $105$~GeV. In addition, it is clear how improved the sensitivity will be through the CTA experiment, for example, fixing the DM mass equal to $1000$~GeV, the current limit given by H.E.S.S. is $\langle \sigma v \rangle \sim 9 \times 10^{-26}$~cm$^3 \,$s$^{-1}$ while the prospects for CTA gives $\langle \sigma v \rangle \sim 4.5 \times 10^{-26}$~cm$^3 \,$s$^{-1}$, in addition, in the less current constrained case, when the DM mass is about $570$~GeV, the current cross section from H.E.S.S. plus Fermi-LAT is $\langle \sigma v \rangle \sim 1.6 \times 10^{-25}$~cm$^3 \,$s$^{-1}$ and for CTA $\langle \sigma v \rangle \sim 4 \times 10^{-26}$~cm$^3 \,$s$^{-1}$, that shows a difference of about one order of magnitude in sensitivity.

In the Figs.~\ref{phihiggslike100} and \ref{phihiggslike500}, we present the constraints for scalar masses equal to $100$~GeV and $500$~GeV, respectively. As already expected, due to the predominant branching fraction of Higgs-like particles into hadrons, the limits follow the same pattern as before, the main difference is in the DM mass interval due to energy conservation.

\subsection{Mediator $\phi + \phi$ decaying predominantly in $e^+e^-$, $\mu^+\mu^-$, $\tau^+\tau^-$ or  $\bar{b}b$}

As the last analysis, we compute the limits for the case where DM annihilates in two scalar fields which decay predominantly in $e^+ e^-$, $\mu^+ \mu^-$, $\tau^+ \tau^-$ or $\bar{b}b$. This study complements the analysis made in \cite{secluded}. 
\begin{figure}[ht]
\centering
\includegraphics[width=0.95\columnwidth]{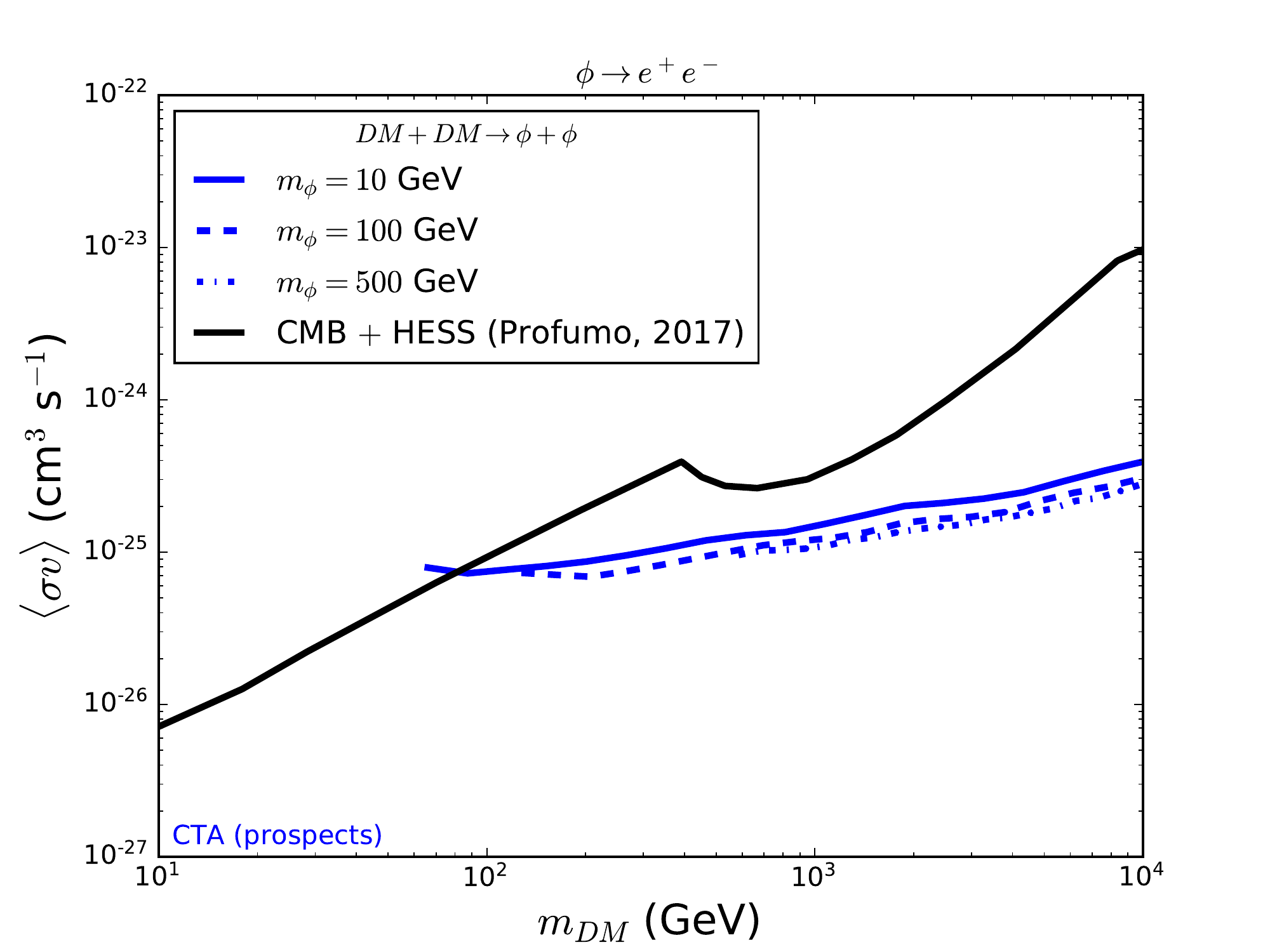}
\caption{Upper limits on the annihilation cross section, with $95\%$~C.L., from the CTA prospects for DM annihilating in the mediator $\phi$ decaying predominantly in $e^+ e^-$ (Blue lines). For comparison, we include the results from \cite{secluded}, see the text for details.}
\label{phiphiee}
\end{figure}
Then, we include here the results from \cite{secluded} (black line) in order to compare with the CTA prospects (blue line). In \cite{secluded}, it was computed the current limits from Fermi-LAT, H.E.S.S., and Planck. It is important to emphasize that we use the results from \cite{secluded}, where $m_{DM} \gg m_{V}$, which is similar and can be compared to our analysis.

\begin{figure}[ht]
\centering
\includegraphics[width=0.95\columnwidth]{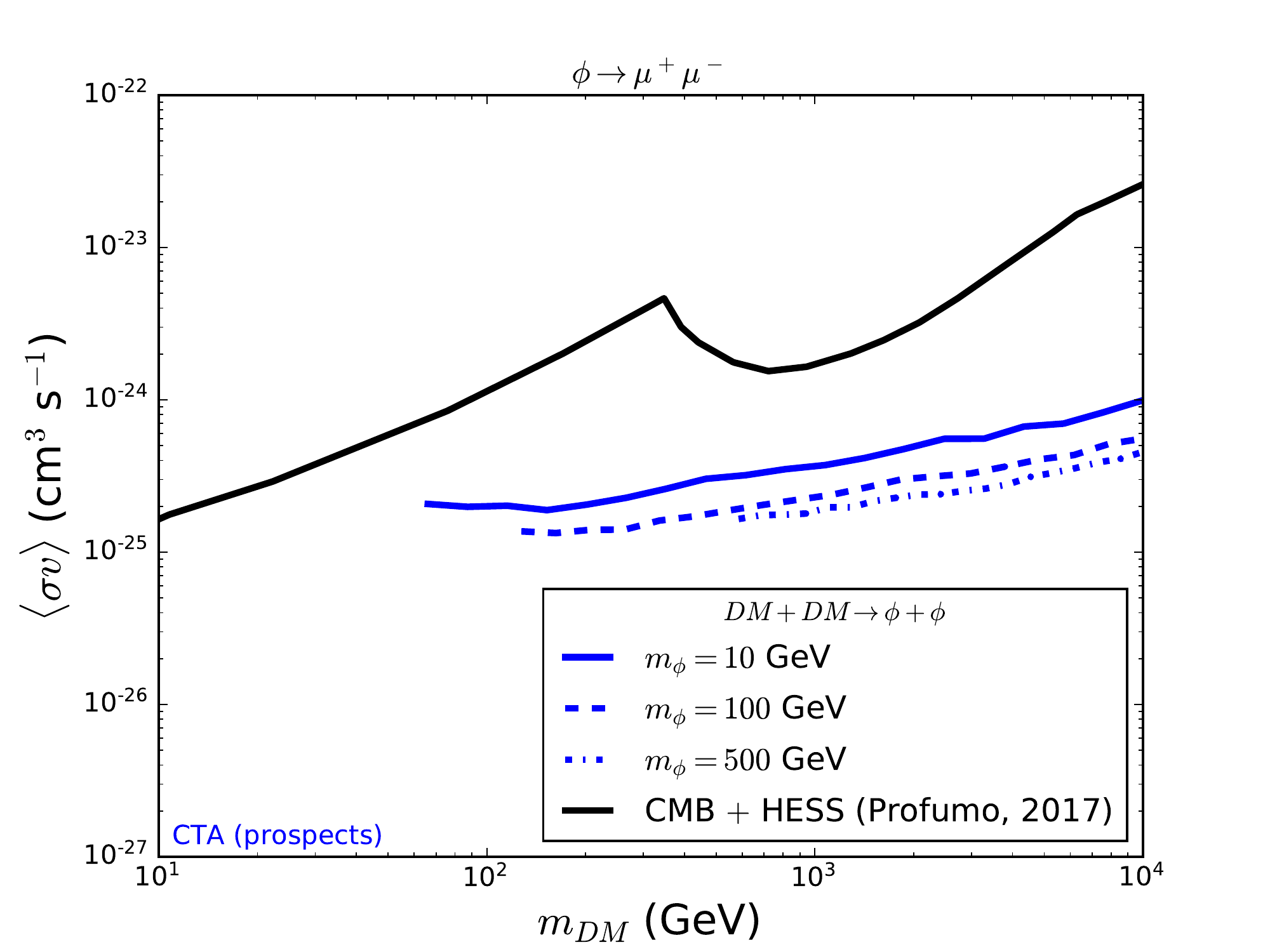}
\caption{Upper limits on the annihilation cross section, with $95\%$~C.L., from the CTA prospects for DM annihilating in the mediator $\phi$ decaying predominantly in $\mu^+ \mu^-$ (Blue lines). For comparison, we include the results from \cite{secluded}, see the text for details.}
\label{phiphimumu}
\end{figure}

In Fig.~\ref{phiphiee}, we show our results for DM annihilating predominantly in $e^+ e^-$, by comparing with the stronger current limits (CMB + H.E.S.S.) \cite{secluded}, we found an improvement in the CTA sensitivity of up to one order of magnitude for high DM masses compared to the current limits, for example, fixing $m_{DM}=5000$~GeV, we have $\langle \sigma v \rangle \sim 2.5 \times 10^{-24}$~cm$^3 \,$s$^{-1}$ for the current limits, while the expected from CTA gives $\langle \sigma v \rangle \sim 1.7 \times 10^{-25}$~cm$^3 \,$s$^{-1}$.
\begin{figure}[ht]
\centering
\includegraphics[width=0.95\columnwidth]{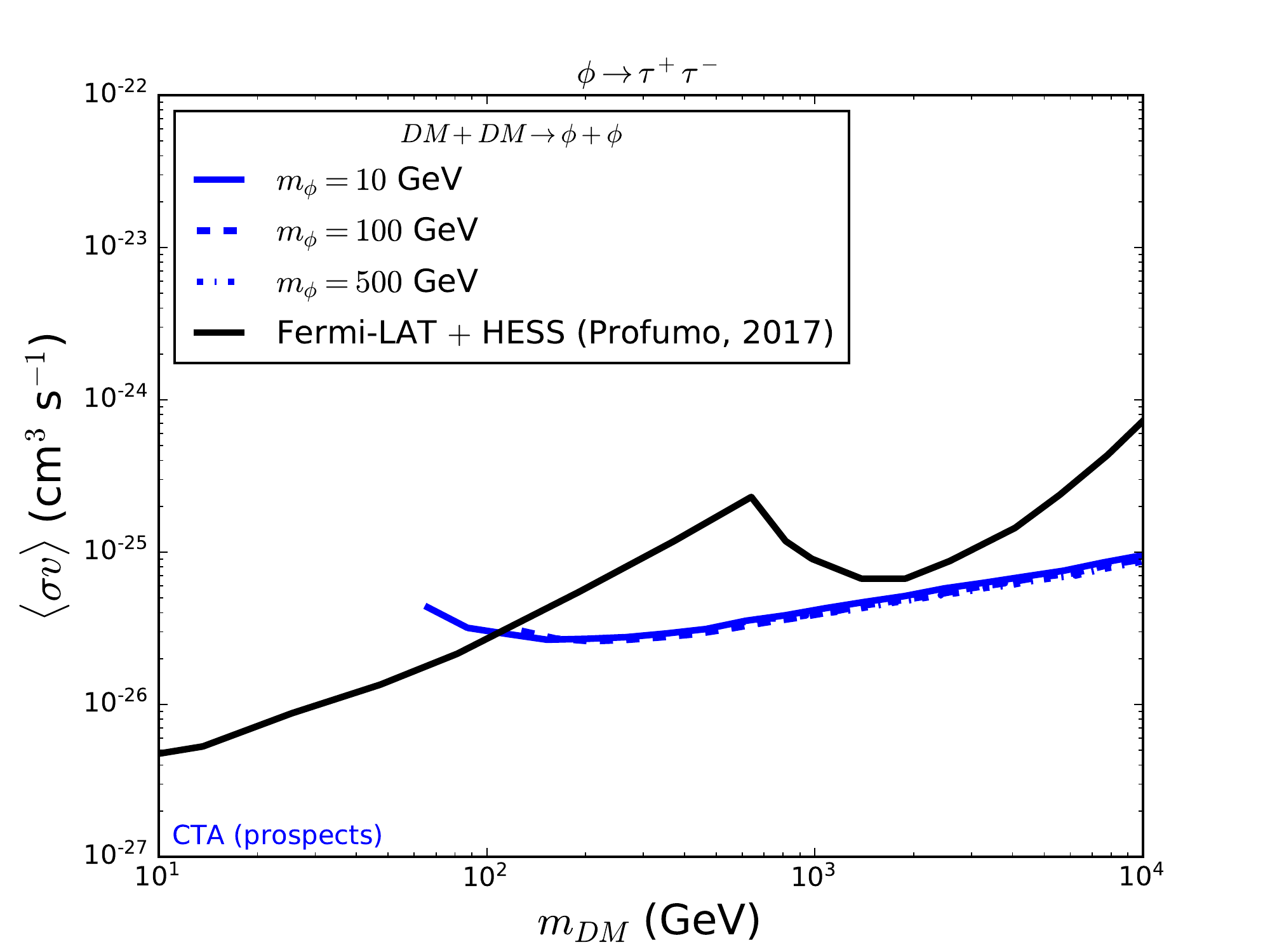}
\caption{Upper limits on the annihilation cross section, with $95\%$~C.L., from the CTA prospects for DM annihilating in the mediator $\phi$ decaying predominantly in $\tau^+ \tau^-$ (Blue lines). For comparison, we include the results from \cite{secluded}, see the text for details.}
\label{phiphitautau}
\end{figure}

In Fig.~\ref{phiphimumu}, we present our results for DM annihilating predominantly in $\mu^+ \mu^-$, by comparing with the current limits (CMB + H.E.S.S.) \cite{secluded}, we found the same gain in sensitivity as before, for example, fixing $m_{DM}=1000$~GeV, we have $\langle \sigma v \rangle \sim 1.6 \times 10^{-24}$~cm$^3 \,$s$^{-1}$ for the current limits, while the expected from CTA gives $\langle \sigma v \rangle \sim 1.8 \times 10^{-25}$~cm$^3 \,$s$^{-1}$ and for $m_{DM}=5000$~GeV we have $\langle \sigma v \rangle \sim 1.1 \times 10^{-23}$~cm$^3 \,$s$^{-1}$ for the current limits, while the expected from CTA gives $\langle \sigma v \rangle \sim 3 \times 10^{-25}$~cm$^3 \,$s$^{-1}$.

It is worth noting that in leptonic final states without hadronic decay, as in the cases mentioned above, the inverse Compton scattering (ICS) can lead to a important contribution to the expected DM flux, especially for high DM masses increasing the sensitivity of the CTA to these channels, we can estimate based on previous analysis \cite{Gomez-Vargas:2013bea} an increase in sensitivity of a factor of a few. We left a detailed analysis for a future work.
\begin{figure}[ht]
\centering
\includegraphics[width=0.95\columnwidth]{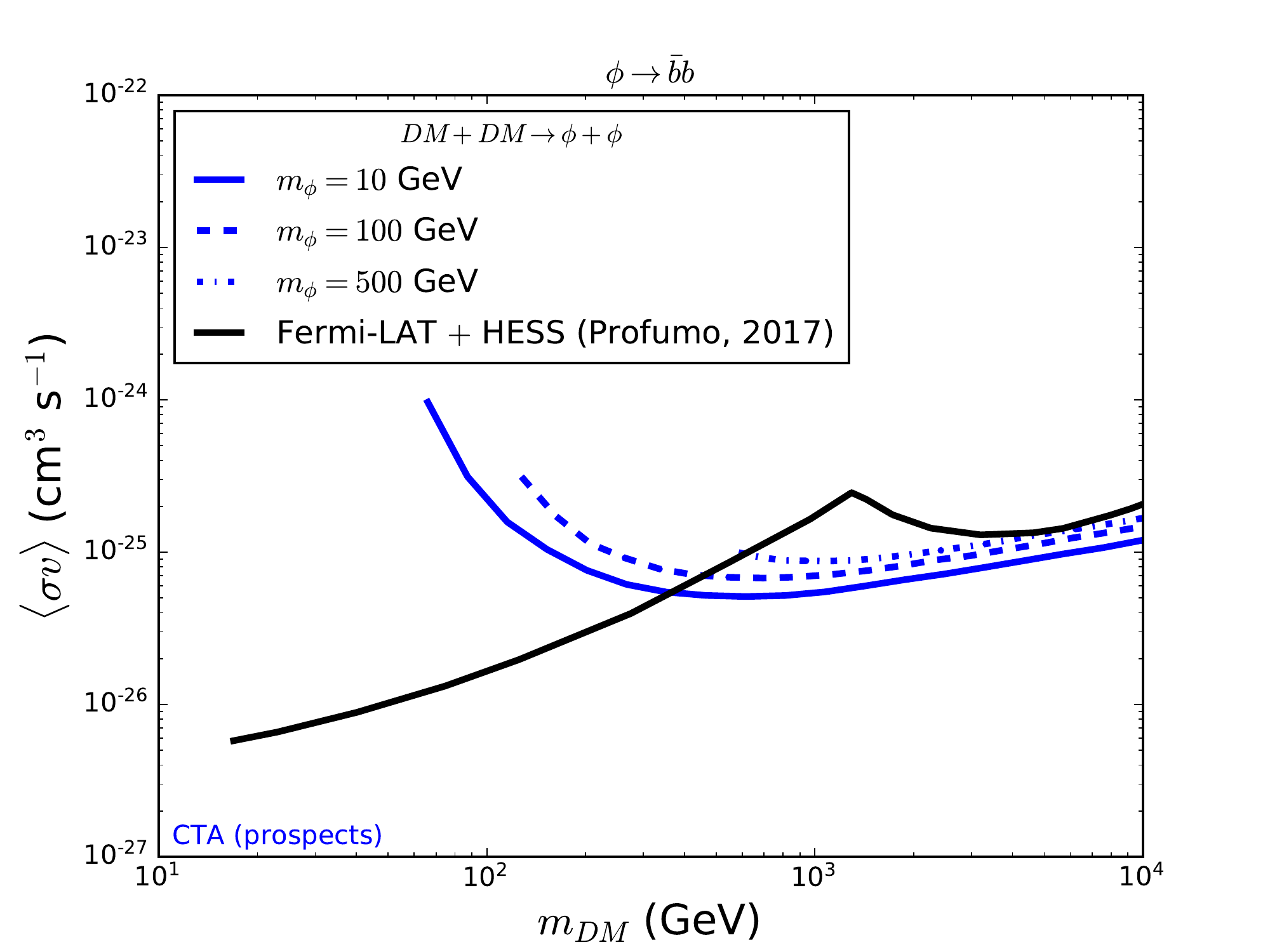}
\caption{Upper limits on the annihilation cross section, with $95\%$~C.L., from the CTA prospects for DM annihilating in the mediator $\phi$ decaying predominantly in $\bar{b}b$ (Blue lines). For comparison, we include the results from \cite{secluded}, see the text for details.}
\label{phiphibb}
\end{figure}

For the $\tau^+\tau^-$ channel (see Fig.~\ref{phiphitautau}), for $m_{DM}=5000$~GeV, the current limit provides $\langle \sigma v \rangle \sim 2.0 \times 10^{-25}$~cm$^3 \,$s$^{-1}$ and for CTA we have $\langle \sigma v \rangle \sim 6.7 \times 10^{-26}$~cm$^3 \,$s$^{-1}$, which is again a huge gain in sensitivity.

In the last case, for $\bar{b}b$ channels (see Fig.~\ref{phiphibb}), the H.E.S.S. and Fermi-LAT limits are strong and the main gain in sensitivity are for DM masses between $600$~GeV and $2000$~GeV. For example, for $m_{DM}=1300$~GeV, the current limit provides $\langle \sigma v \rangle \sim 2.5 \times 10^{-25}$~cm$^3 \,$s$^{-1}$ and while for CTA $\langle \sigma v \rangle \sim 6.3 \times 10^{-26}$~cm$^3 \,$s$^{-1}$.

\begin{figure}[ht]
\centering
\includegraphics[width=0.95\columnwidth]{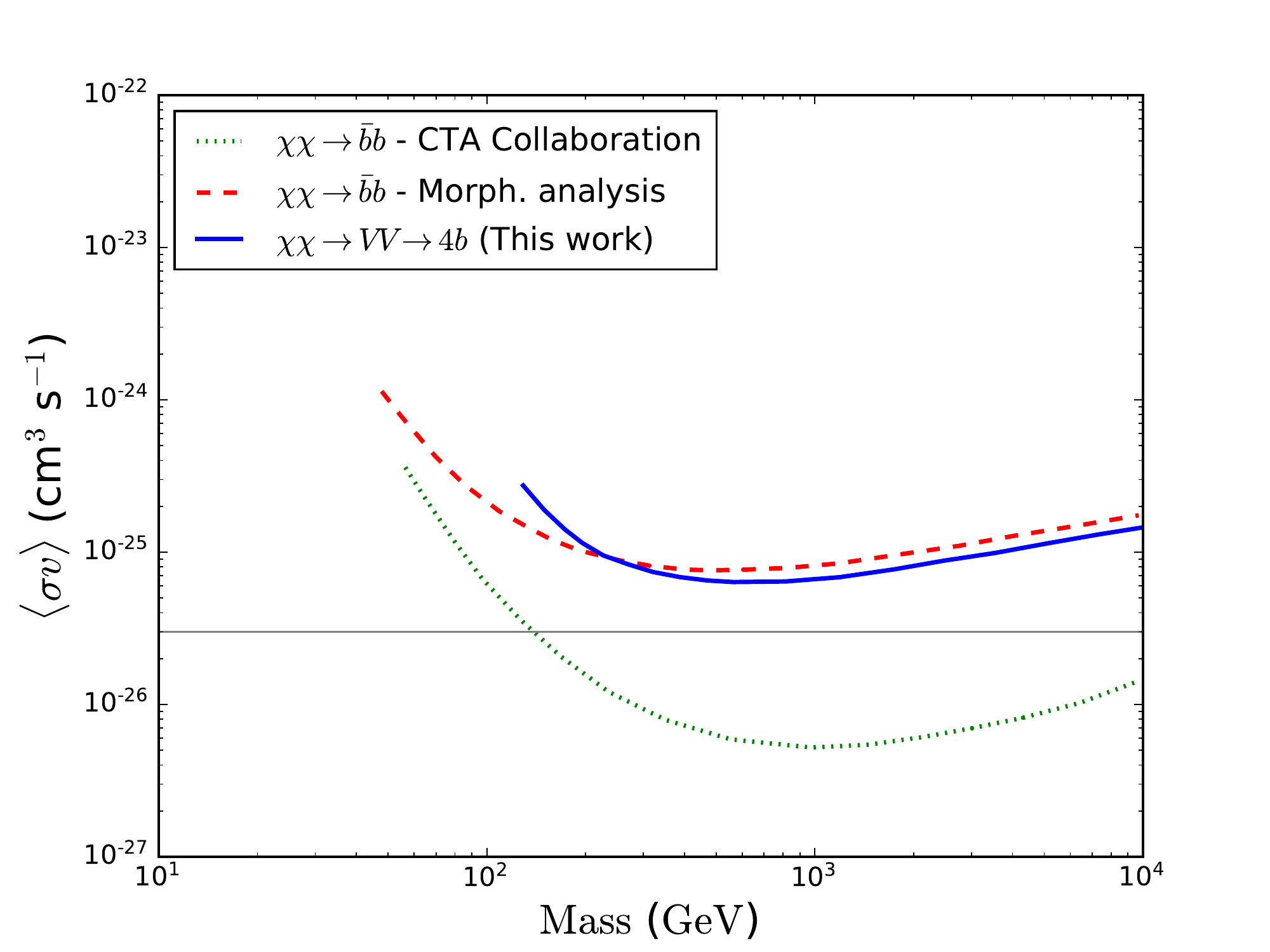}
\caption{Comparison between different analysis for the CTA limits compared with our limits for $100$~h of observation. Blue line, our limit for $\phi \rightarrow \bar{b}b$ with $m_\phi = 100$~GeV. Red dashed line, limit using the same method used here for the $\bar{b}b$ final state. Green dotted lines, the limits for the same channel reported by the collaboration.  }
\label{comparison}
\end{figure}

For completeness, we include a comparison between different analysis using CTA prospects (see Fig. \ref{comparison}), in these cases the profile chosen was the Einasto, and we include our limits for $\phi \rightarrow \bar{b}b$ with $m_\phi = 100$~GeV (blue line), and the limits for $\bar{b}b$ using the same morphological analysis used in this work (red dashed line), a conservative scenario \cite{Silverwood:2014yza}, and the limits for the same channel provided by the CTA collaboration (green dotted line). As we expect, since the secluded scenarios provides a harder spectrum, the limit is slightly stronger than the direct annihilation, especially because of the high sensitivity to higher energies expected for CTA \cite{Morselli:2017ree}. The limits reported by the collaboration do not take into account GDE or systematic uncertainties, so it is very optimistic, as mentioned before.\\

Summarizing, the limits from CTA over some secluded models show an improvement in sensitivity compared to the current data, however, the prediction for CTA do not show a huge increment in sensitivity compared to H.E.S.S., this is due to two main facts: firstly, the chosen regions of interest ON and OFF are completely different, the other point can be related to the GDE, because the H.E.S.S. experiment do not take it into account \cite{Silverwood:2014yza}. Anyway, the CTA prospects for secluded models can reach up to one order of magnitude in sensitivity depending on the channel compared to the strongest current limits, like Fermi-LAT, H.E.S.S. and Planck.\\

\section{Conclusions}
\label{sec:con}

In this work we computed the gamma-ray fluxes for different secluded scenarios, where we include a DM particle annihilating in two scalars, with three different masses $10$~GeV, $100$~GeV and $500$~GeV, in order to compare the current, Fermi-LAT, H.E.S.S. and Planck, with the future limits expected from the CTA experiment.

In the first case, we choose the mediator as a Higgs-like scalar, with the branching ratio changing accordingly, in this scenario we computed the gamma-ray spectrum and flux, and obtained the upper limits on the annihilation cross section using data from Fermi-LAT, looking at dSphs, from H.E.S.S. looking at the galactic centre, from Planck via CMB searches and the prospects from CTA experiment looking at the GC. Here, we excluded the canonical annihilation cross section $\langle \sigma v \rangle = 3 \times 10^{-26}$~cm$^3 \,$s$^{-1}$ for DM masses below $105$~GeV. In addition, we showed that when the DM mass is about $570$~GeV, the current limit on the cross section from H.E.S.S. plus Fermi-LAT is given by $\langle \sigma v \rangle \sim 1.6 \times 10^{-25}$~cm$^3 \,$s$^{-1}$, while the expected sensitivity for CTA gives $\langle \sigma v \rangle \sim 4 \times 10^{-26}$~cm$^3 \,$s$^{-1}$, that shows a difference of about one order of magnitude in sensitivity. 

In the second analysis, we computed the upper limits on the annihilation cross section using the prospects for the CTA experiment for the mediators as a leptophilic or a leptophobic scalar, and we saw how the CTA will increase the sensitivity to secluded models in the next years. For example, for $\phi$ decaying predominantly in $e^+e^-$ with $m_{DM}=5000$~GeV, we have $\langle \sigma v \rangle \sim 2.5 \times 10^{-24}$~cm$^3 \,$s$^{-1}$ for the current limits, while the expected from CTA gives $\langle \sigma v \rangle \sim 1.7 \times 10^{-25}$~cm$^3 \,$s$^{-1}$, showing again a difference of about one order of magnitude in sensitivity.

Besides, we found that independently of the channel, the CTA experiment will increase the sensitivity compared to the current experiments for DM masses above $400$~GeV, increasing the current bounds by up to approximately one order of magnitude in these secluded scenarios.
\section*{Acknowledgments}

CS acknowledges Farinaldo Queiroz, Ma\'ira Dutra and Antonio Santos for useful comments and discussions in reviewing the manuscript. CS is also grateful to the referee for useful questions and comments about the paper. This work was supported by MEC and UFRN.

\bibliographystyle{JHEPfixed}
\bibliography{darkmatter}

\end{document}